\documentclass[aps,prb,reprint,showpacs,superscriptaddress,groupedaddress]{revtex4-1}
\usepackage{graphicx}
\usepackage{amsmath}
\usepackage{amssymb}
\usepackage{dcolumn}
\usepackage{latexsym}
\usepackage{rotating}
\usepackage[usenames,dvipsnames]{xcolor}  
\usepackage{float}
\usepackage{epsfig}
\usepackage{psfrag}
\usepackage{natbib}
\usepackage{bm}
\usepackage{eucal}
\usepackage{braket}
\usepackage{enumerate}
\usepackage{longtable}
\usepackage{subfigure}
\usepackage{bm}
\usepackage{hyperref}
\usepackage{amsfonts}
\usepackage{dcolumn}
\usepackage{bm}
\usepackage{subfigure}

\newcommand{\be}{\begin{equation}}
\newcommand{\ee}{\end{equation}}
\newcommand{\bn}{\begin{eqnarray}}
\newcommand{\en}{\end{eqnarray}}

\usepackage{color} 

\usepackage{hyperref}
\begin{document}

\author{S. Koley$^{3}$}\email{sudiptakoley20@gmail.com}
\author{M. S. Laad$^{1}$}\email{mslaad@imsc.res.in}
\author{A. Taraphder$^{2}$}\email{arghya@phy.iitkgp.ernet.in} 

\title{Do Dynamical Excitonic Liquid Correlations Mediate Superconductivity in Elemental Bismuth? }
\affiliation{$^{1}$Instt. Math. Sciences Taramani, Chennai, 600113 and 
Homi Bhabha National Institute, India}
\affiliation{$^{2}$Department of Physics and Centre for Theoretical Studies, 
Indian Institute of Technology, Kharagpur, 721302 India}
\affiliation{$^{3}$ Department of Physics, St. Anthony's College, Shillong,
Meghalaya, 793001 India}
\begin{abstract}    
\noindent Motivated by the remarkable discovery of superconductivity in elemental Bismuth,
we study its normal state in detail using a combination of 
tight-binding (TB) band-structural supplemented by dynamical mean-field theory (DMFT). We show that a two-fluid model composed of preformed and dynamically fluctuating excitons coupled to a tiny number of carriers provides a unified rationalization of a range of ill-understood normal state spectral and transport data.  Based on these, we propose that resonant scattering involving a very low density of renormalized carriers and the excitonic liquid drives logarithmic enhancement of vertex corrections, boosting superconductivity in $Bi$.  A confirmatory test for our 
proposal would be the experimental verification of an excitonic semiconductor with electronic nematicity as a `competing order' on inducing a semi-metal-to semiconductor transition in $Bi$ by an external perturbation like pressure.

\end{abstract}
\maketitle
\noindent      Rhombohedral Bismuth (Bi) has recently acquired prominence in a variety 
of contexts.  Like graphite, elemental Bi also shows a 
magnetic field- as well as a pressure-induced metal-insulator-like 
transition~\cite{Bi-MIT} with a large magneto-resistance, and electron 
fractionalization in high magnetic fields~\cite{behnia}.  Remarkable discovery 
of superconductivity (SC) in Bi at very low temperature 
($T_{c}\simeq 0.5$~mK)~\cite{ramky} in a lowest carrier elemental system to date adds to the range of novel behaviors exhibited by this `simple' system.  Experiments conclusively establish that SC in Bi is {\it not} of the standard 
BCS variety and, in fact, it would be more aptly characterized as being in a non-adiabatic limit of pairing theory if electron-phonon coupling were to be invoked as the dominant pairing glue.

\noindent    The unique electronic properties of $Bi$ arise from successive distortions 
of a higher-symmetry simple-cubic structure.  First, a small relative 
displacement along the body diagonal doubles the unit cell.  By itself, this 
would drive $Bi$ into a Peierls-like band semiconductor.  But additional 
rhombohedral shear causes further lowering of symmetry, allowing valence and 
conduction band overlap - in good accord with band structure studies, 
and explains why $Bi$ is metallic.  Due to extremely small Fermi pockets 
(of size $10^{-5}$ of the Brillouin zone) and tiny carrier density 
($3\times 10^{17}$cm$^{-3}$), the long mean-free path accounts for its small low-$T$ resistivity.  One might then think that traditional one-electron band structure is an adequate framework to understand its electronic properties.

\noindent    However, careful perusal of extant data points toward a much more 
interesting situation.  Early data~\cite{kukkonen-J.Phys.F-1977} show that the electrical resistivity $\rho_{dc}(T)\simeq \rho_{0}+AT^{2}$ for $T < 50$~K.  But very unusual
behavior at low $T$, with $\rho_{dc}(T)\simeq T^{5}$ was recently seen that invoked a plasmaron picture~\cite{giamarchi}.  Optical data raise additional issues in this context~\cite{armitage}.  While a tiny Drude 
component is visible at low energy, finding of sizable mid-infra-red (mid-IR) 
absorption is inexplicable in a free-electron framework.  Given the tiny Fermi energy $\simeq 25$~meV, a large $T$-dependent transfer of spectral weight over a much larger scale $\simeq 300$~meV also presents a challenge for the standard 
(uncorrelated band) view. On the theoretical front, first-principles density-functional theory (DFT) and Slater-Koster tight-binding fits~\cite{ting1994} 
conclusively show that: $(i)$ there are multiple tiny pockets in $Bi$, with 
two being almost perfectly compensated, and the others supplying a tiny 
additional number of carriers. Moreover, the importance of spin-orbit coupling 
(SOC) is shown by the fact that the correct positions of the e(h) pockets are 
only found when SOC is included in band calculations.
While DFT+SOC calculations indeed yield the correct shape and size of the carrier pockets, they cannot, by construction, rationalize the above features.  It is long known~\cite{armitage} that the dominant interaction between conduction and valence band carriers in $Bi$ is of short-range excitonic character, but its role in $Bi$ has never been satisfactorily addressed. Generically, one also expects a symmetry-adapted coupling of such interband exciton-like entities to intervalley phonons~\cite{cohen}. As far as SC is concerned, the large discrepancy~\cite{ramky} between the measured and calculated (within BCS theory) ratio of the upper critical field to $T_{c}$ also indicates a non-adiabatic `strong coupling' SC.
  These limited observations must constrain theoretical modelling: both unconventional metallicity and SC must find explication in a picture based on $(i)$ the special band structure of Bi and $(ii)$ strong scattering (electron-hole and/or electron-phonon) processes beyond DFT and, in particular, microscopic processes which generate a $\rho_{dc}\simeq T^{5}$ should also be involved in generating the SC pair glue.

\noindent Our first observation is that two of the multiple electron(hole) pockets of Bi are almost perfectly compensated, leading to a situation famously encountered in transition-metal dichalcogenides (TMD), where preformed excitonic liquid (PEL) driven charge-density-wave (CDW) states are ubiquitous~\cite{2H-TaSe2,1T-TiSe2}. Band calculations reveal a total of four pockets in $Bi$, of which two (electron and hole) exhibit a propensity for an excitonic instability. Were there only these two pockets, one would have expected an electron-hole attraction-mediated excitonic insulator.  However, the remaining pockets now supply a tiny number of additional carriers, leading to a physical picture of {\it an incipient excitonic insulator self-doped with a tiny number of carriers}.  Implicit herein is the fact that poor screening of the interband Coulomb interaction due to ultra-low carrier density makes this a strongly interacting system. Thus, we are led to a model of a self-doped excitonic insulator (EI) in the intermediate coupling limit, where novel physics can arise from strong scattering between carriers and preformed but uncondensed excitons. In this letter, we establish this using  tight-binding-plus dynamical mean-field theory (TB-DMFT) calculations as done earlier~\cite{2H-TaSe2,1T-TiSe2}.  Having good accord with normal state transport, we elucidate a `strong coupling' electronic mechanism, wherein strong resonant scattering between carriers and preformed excitons enhances the SC $T_{c}$.

\begin{figure}
\centering
(a)
\includegraphics[angle=270,width=0.8\columnwidth]{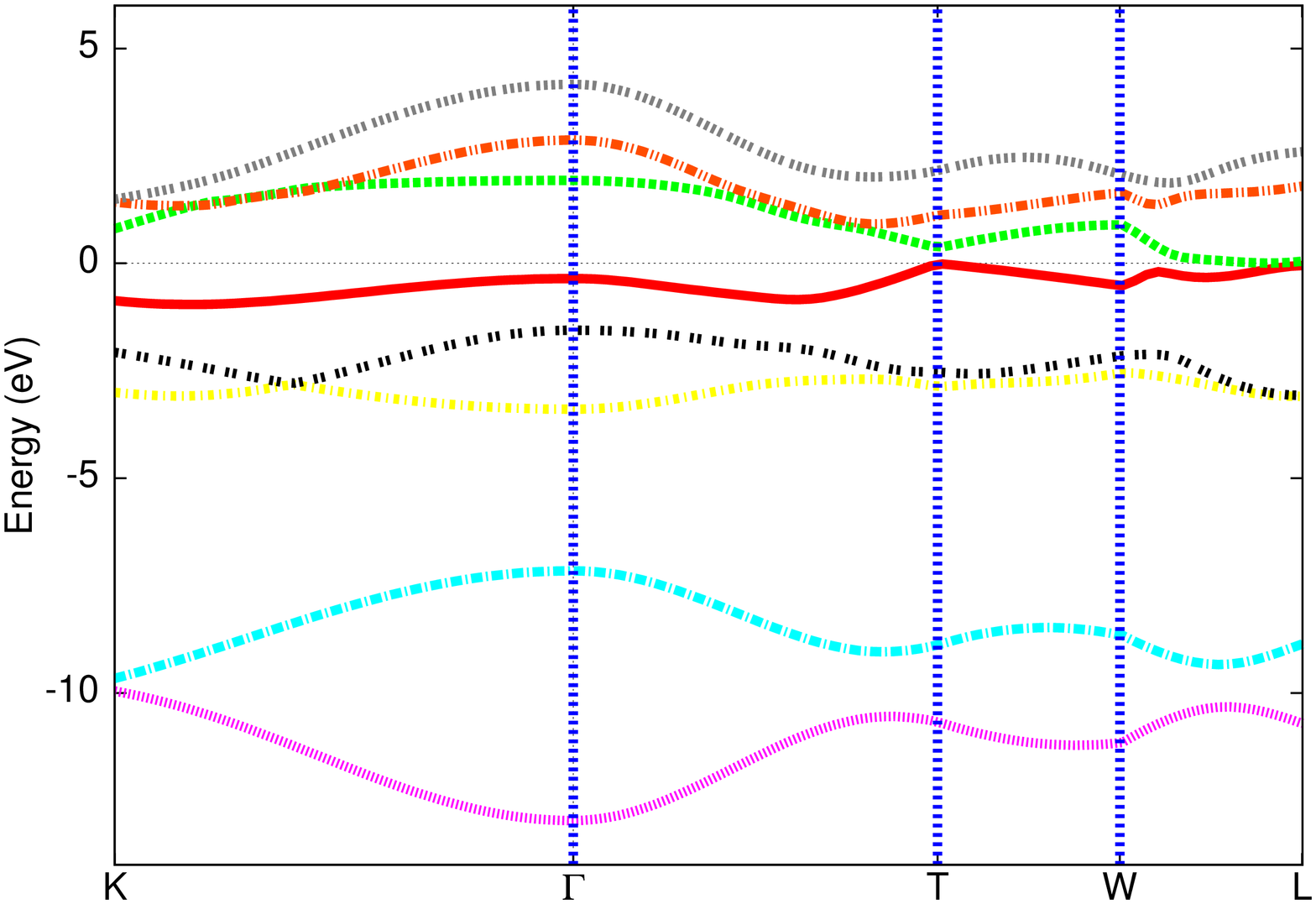}
\caption{(Color Online)Tight binding (TB) bands of $Bi$ in the rhombohedral structure including spin-orbit coupling (SOC) as done in Ref.~\cite{ting1994}.  For the effective two-band model incorporating excitonic correlations within DMFT calculations, we use only the two (red and green) bands crossing the Fermi 
energy, $E_{F}(=0)$.}
\label{fig1}
\end{figure}

\noindent We begin by using the Slater-Koster (SK) fit with form factors and parameters {\it including} SOC as in earlier
work~\cite{ting1994}. The resulting band structure in Fig.~\ref{fig1} excellently reproduces all the pockets seen in 
full DFT calculations, constituting the appropriate band structural input for the correlation calculations. Guided by the discussion above, we focus on inter-band excitonic correlations.  The two-band Hubbard model incorporating the two bands crossing the Fermi energy is  $H=H_{0}+H_{1}$, with
\be
H_{0}=\sum_{(\mu,\nu=1,2),k,\sigma}\epsilon_{\mu}(k)c_{k\mu\sigma}^{\dag}c_{k\nu\sigma}+h.c) + \Delta\sum_{i}(n_{i,1}-n_{i,2})
\ee
where $\epsilon_{1}(k)=E_{p}+3V_{pp\sigma}$cos$(\frac{\sqrt{3}k_{x}}{2})$cos$(\frac{k_{y}}{2})+V_{pp\pi}[$cos$(\frac{\sqrt{3}k_{x}}{2})$cos$(\frac{k_{y}}{2})+2$cos$k_{y}]$ (represnted by the green line in 
Fig. ~\ref{fig1}), $\epsilon_{2}(k)=E_{p}+V_{pp\sigma}[$cos$(\frac{\sqrt{3}k_{x}}{2})$cos$(\frac{k_{y}}{2})+2$cos$k_{y}]+3V_{pp\pi}$cos$(\frac{\sqrt{3}k_{x}}{2})$cos$(\frac{k_{y}}{2})$ (red line in Fig.~\ref{fig1}) where E$_p$ is on-site energy, V$_{pp\pi}$ and V$_{pp\sigma}$ are third nearest neighbour interaction and the interband matrix element $V_{12}(k)\simeq 2i\sqrt{3}$sin$(\frac{\sqrt{3}k_{x}}{2})$sin$(\frac{k_{y}}{2})$, as extracted earlier~\cite{ting1994} from an SK fit. Here $\mu, \nu$ represent the band indices for the two bands crossing $E_{F}(=0)$ in Fig.~\ref{fig1}. The local terms are 
\be
H_{1}=\sum_{\mu=1,2}U_{\mu\mu}\sum_{i}n_{i\mu\uparrow}n_{i\mu\downarrow} + U'\sum_{i}n_{i,1}n_{i,2}
\ee
 Since $g=\omega_{D}/E_{F}\simeq 0.5$ ($g$=electron-phonon coupling, $E_{F}$=Fermi energy) is actually sizable in 
Bi~\cite{ramky}, one is in a non-adiabatic limit.  In contrast to the anti-adiabatic ($g\rightarrow\infty$) or the adiabatic ($g\rightarrow 0$) limits, one cannot `integrate out' the (intervalley in Bi) phonons to give a further e-h attraction $O(g^{2}/\hbar\omega_{D})$.  One must solve $H$ including an explicit $e-p$ term 
by coupling the intervalley phonons to the ${\bf k}$-dependent hybridization as done in earlier work~\cite{1T-TiSe2}.
\vspace{1.0cm}

\noindent {\bf TB+DMFT Results and Transport}

\noindent We solve the two-band model above using DMFT, with multi-orbital IPT as the 
`impurity' solver. Though not numerically exact, it works very well in 
real multi-band cases where there is a considerable crystal-field splitting 
between the bands (here, between valence and conduction band). It is a fast
solver and its efficacy in a wide range of real systems is 
known~\cite{1T-TiSe2,jarrell}. In contrast, though QMC solvers are much more reliable, they cannot access temperatures below $O(20)$~K at present, which makes them unsuitable for investigation of the low $T(<10$~K) states. We choose $U=0.5-0.7$~eV and $U'=0.1-0.2$~eV as appropriate parameters, and while ab-initio estimates will yield more precise estimates, our present choice is physically motivated. Given widths 
of $O(1.0-1.5)$~eV for the valence (VB) and conduction (CB) bands, we are in the intermediate coupling limit of the two-band model. This is precisely the case where DMFT works best (for specific studies in the BCS-BEC crossover, see~\cite{U<0Hubbardmodel}).

\begin{figure}
\includegraphics[angle=270,width=\columnwidth]{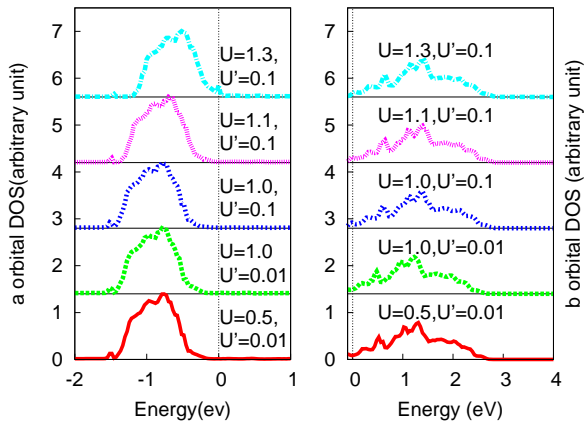}
\caption{(Color Online) The orbitally resolved density of states from TB+DMFT at various U and U$'$.}
\label{fig2}
\end{figure}

\begin{figure}
\includegraphics[angle=270,width=\columnwidth]{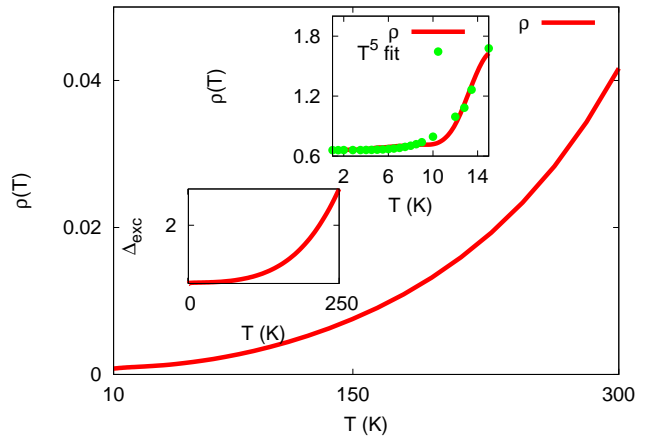}
\caption{(Color Online) $DC$ Resistivity for $Bi$ within TB+DMFT: a $T$-dependent crossover from $\rho_{dc}(T)\simeq T$ at `high'-$T$ to a
correlated Landau-Fermi liquid like $\rho_{dc}(T)\simeq T^{2}$ below $T\simeq 90$~K is followed by $\rho_{dc}(T)\simeq T^{5}$ form at very low $T$ (see inset for comparison with data points) is clear.  The lower-left inset shows the $T$-dependent {\it local} interband excitonic average: onset of $\rho_{dc}(T)\simeq T^{5}$ correlates with a decreasing excitonic average. The y-axis of both the 
insets are divided by 10$^{-3}$.}
\label{fig3}
\end{figure}

\noindent In Fig.~\ref{fig2}, we show the TB+DMFT local density-of-states (LDOS) as $U,U'$ are cranked up.  While correlations gradually close the band gap for the VB (`$a$'-orbital) as expected~\cite{mohit}, they reduce the LDOS at $E_{F}$ for the CB (`$b$' orbital): these contrasting behaviors are direct 
consequences of distinct effects of local electronic correlations on band-insulating and metallic subsets of the non-interacting band structure. Around $U\simeq W_{b}$, the $b$-fermion band-width, we observe eventual opening of a {\it Mott}-like gap in the LDOS, seen by the fact that Im$\Sigma_{b}(\omega)$ 
develops a pole structure at $E_{F}$ in this case.  Correspondingly, the $a$-fermion states retain metallic character: this would be an interesting manifestation of orbital-selective Mott physics in semi-metals and, were such a strong coupling regime to be realized, would open up the possibility to 
development of novel instabilities with concomitant competing orders~\cite{laad-122} as, for instance, in Fe-arsenides.  However, this is {\it not} the regime applicable to $Bi$, and so we concentrate on the smaller $U$ regime.
Choosing $U=0.5$~eV, $U'=0.15$~eV, we next compute the $dc$ resistivity using the Kubo formalism in DMFT.  This task is facilitated by the finding~\cite{biermann} that irreducible vertex corrections 
apprearing the Bethe-Salpeter equations for conductivities are negligible and can be ignored to a very good approximation.  Interestingly, as shown in Fig.~\ref{fig3}, we find that a linear-in-$T$ behavior of 
$\rho_{dc}(T)$ at `high' $T\geq 40-50$~K smoothly crosses over to a Fermi-liquid-like $T^{2}$ behavior up to about $10$~K and, remarkably, exhibits a further low-$T$ crossover to a `good' metal with 
$\rho_{dc}(T)\simeq T^{5}$.  This is in very good accord with experimental trends, and mandates deeper microscopic rationalization.  To cement the link between 
transport and excitonic liquid fluctuations, Fig.~\ref{fig3} shows the {\it excitonic} average, computed as $\Delta_{exc}=(-1/\pi)\int d\omega$Im$G_{12}(\omega)$ 
($G_{12}(\omega)$ is itself computed from two-band DMFT(IPT) as previously done for TMDs~\cite{1T-TiSe2}).  We find a clear correlation between $\Delta_{exc}(T)$ and the $T$-dependence of 
$\rho_{dc}(T)$ over a wide $T$ range.  This provides strong theoretical evidence linking (at least) the $dc$ resistivity to microscopic processes involving scattering of the tiny number of carriers off well-formed and
quasi-local excitonic correlations: at high-$T$, the latter are incoherent, leading to a quasi-linear-in-$T$ resistivity, while increasing one-fermion coherence via suppression of incoherent 
excitonic fluctuations provides a rationalization for the near $T^{2}$ at intermediate $T$.  Remarkably, however, we also obtain the $\rho_{dc}(T)\simeq T^{5}$ dependence at very low $T$, the latter correlated with a reduced excitonic fluctuation at low $T$.
  Thus, we identify a new element: the $T$-dependence of sizable and dynamical 
`preformed' excitonic correlations governs the $T$-dependent resistivity in a wide $T$ window.  Our 
proposal is distinct from: $(i)$ the plasmaron view~\cite{giamarchi}, where weak-coupling RPA-like analysis, based upon a picture of efficient screening, is employed to get $\rho_{dc}\simeq T^{5}$ from long-range interactions, and $(ii)$ pure e-p coupling models, which could, in principle, also yield similar behavior when $T<<\Theta_{D}$, something that may also obtain in $Bi$.  We propose that a way to distinguish between 
these distinct scenarios could be $T$-dependent tunnelling measurements at small-to-intermediate $T$: 
as a function of $T$, the conductance $g(V)=dI/dV$ would, in a picture involving coupling of carriers to 
any bosonic mode(s), show finite-voltage (energy) peak-dip-hump features. The energies and spectral weights 
of such features could be compared with estimates from different bosonic channels, allowing a determination of the most important fermion-boson scattering channel. However, since there will always be a symmetry-dictated 
coupling of interband excitons to intervalley phonons, one also generally 
expects {\it two} bosonic modes at different energies to show up in $g(V)$: 
their relative weights provide an estimate of the relative importance of 
carrier-exciton vis-a-vis carrier-phonon coupling.
   
\begin{figure}
\includegraphics[angle=270,width=\columnwidth]{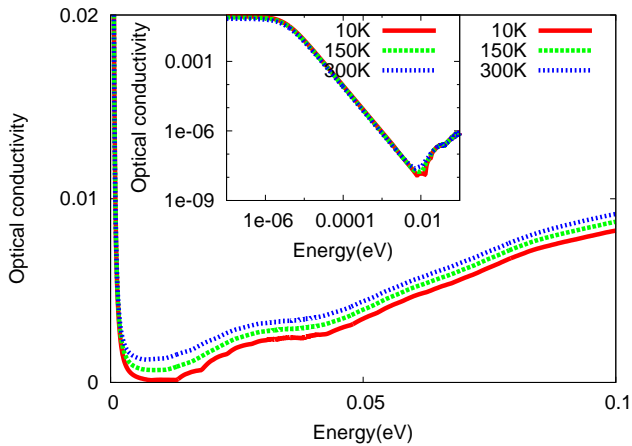}
\caption{(Color Online) TB+DMFT Optical Conductivity in the normal state for $Bi$.  A tiny Drude-like contribution is followed by inter-band absorption: this feature onsets at $\simeq 15$~meV and originates from strong resonant scattering between the tiny number of carriers and interband excitons in DMFT.  The pre-peak in the absorption, centered at $\simeq 10$~meV, also originates from the same mechanism.  The inset reveals the same features in more detail on a log-log plot.  These findings are in very good accord with data~\cite{armitage}}
\label{fig4}
\end{figure}

\noindent   Further support for our view arises when we compare the DMFT optical conductivity, $\sigma_{xx}(\omega)$ in Fig.~\ref{fig4} as a function of $T$ with 
published data~\cite{armitage}.  Since we have kept only the two lowest bands crossing $E_{F}$ from the SK fit, we do not expect accord at higher energies, but can 
readily make a comparison for the relevant energy range (few hundred milli-eV) of interest.  Specifically, up to about $100$~meV, our result matches quite well 
with data, including $(i)$ the plasmon edge, $(ii)$ the detailed optical lineshape as a function of energy up to about $100$~meV, $(iii)$ sizable optical spectral weight transfer upon raising $T$.  The 
plasmon-like features are clearly visible on both regular and log-log plots as a clear absorption onset at $\simeq 15$~meV at low $T=10$~K.  Interestingly, it is also 
{\it preceded} by a `prepeak' structure, centered at $\simeq 10$~meV, in good accord with observations.  This prepeak feature is also washed out with increasing $T$, in accord with data.  Clear spectral weight transfer up to $\simeq 200$~meV is also found: given the tiny $E_{F}$ in $Bi$, this is quite a large energy scale, attesting to considerable dynamical correlations. Finally, we find an isosbestic point around $5.0$~meV as a function of $T$, which is another characteristic signature of dynamical electronic correlations that could be tested in extant work.  

\noindent    Taken together, our results strongly support the idea that a common underlying scattering process involving tiny number of carriers and uncondensed fluctuating interband 
excitons may be operative in $Bi$.  That the dominant interaction in $Bi$ has a short-range excitonic character is well known~\cite{armitage}.  Our results show that 
it seems to be a sufficient {\it minimal} input to understand transport in $Bi$, and ties in our view with plasmarons~\cite{giamarchi}).  Generally, mid-infra-red features in correlated metals ($e.g$, DMFT for one-band Hubbard type models) involve interband transitions between low-energy itinerant and high-energy localized states. In our case, they arise from transitions involving carrier states 
scattering off the incoherent and interband excitons involving precisely the renormalized VB and CB, since $\sigma_{xx}(\omega)$ in DMFT is just a direct convolution of the one-particle DMFT spectral functions.  In $Bi$, this feature occurs at a very low energy because of the tiny $E_{F}$.  The interband electronic excitations involving carriers coupled to such exciton-like entities have been christened {\it plasmaron} in prior work.  We find that this characterization is not in conflict with our excitonic fluctuation picture, since we have now shown that it can be well described by analyzing effects of predominantly local and sizable excitonic liquid-like correlations in a quasi-realistic model for $Bi$.  

\noindent Buoyed by the agreement with the phenomenology so far, we venture to propose a specific model for superconductivity in $Bi$. To do so, we need to generate an effective pairing interaction
arising from excitonic and/or phonon fluctuation exchange.  Given the efficacy of the excitonic liquid view above for the normal state, and that the exciton formation
energy scale ($\Omega_{ex}\simeq \Delta_{exc}$) is quite large, we are clearly in the non-adiabatic limit.  This precludes a BCS-like instability to SC in $Bi$.  In the normal state, the interband excitons involving the VB and CB form an isospin${\bf T}=1/2$ degree of freedom, described by $T^{+}=\sum_{k,\sigma}a_{k\sigma}^{\dag}b_{k\sigma}, T^{-}=\sum_{\sigma}b_{k\sigma}^{\dag}a_{k\sigma}, T^{z}=\sum_{k}(n_{ka}-n_{kb})/2$ in particle-hole space.  Precisely as in the $U<0$ Anderson lattice~\cite{CKL}, one has a crossover to an incoherent exciton fluctuation dominated regime at $k_{B}T\simeq O(U')$, followed by a second crossover to a quasi-coherent exciton fluctuation dominated regime at a lower exciton-Kondo scale, $T_{K}^{ex}\simeq U'(\pi J\rho)^{1/2}e^{-1/(2J\rho)}$, where $\rho=\rho(E_{F})$ is the LDOS at $E_{F}$ and $J\simeq t_{12}^{2}\chi_{12}(0)$, with $\chi_{12}(0)$ the excitonic susceptibility at $\omega=E_{F}(=0)$: correlated FL behavior obtains below $T_{K}^{ex}$, thanks to an excitonic Kondo screening implicit in a situation where a tiny number of carriers are now coupled to fluctuating isospins ${\bf T}$.  It is precisely this resonant scattering that governs transport in the normal state in good accord with transport data as found above.

\noindent As $T$ is lowered, intersite, residual interactions develop in full analogy with the Kondo lattice.  They correspond to exchange coupling between isospins, and can be viewed as an exciton fluctuation-exchange
process. The residual interactions read

\be
H_{eff}^{(2)}\simeq J\sum_{<i,j>,\sigma,\sigma'}a_{i\sigma}^{\dag}b_{j\sigma}b_{j\sigma'}^{\dag}a_{i\sigma'}
\ee
which is also $J\sum_{<i,j>,\sigma,\sigma'}[n_{ia\sigma}-a_{i\sigma}^{\dag}b_{j,-\sigma}^{\dag}{\dag}b_{j\sigma}a_{i,-\sigma}]$.  Absorbing the first term in the normal state Hamiltonian, $H_{res}^{(2)}$ in momentum space is 
$$H_{res}^{(2)}=-J\sum_{kpq}(a_{k\sigma}^{\dag}b_{p,-\sigma}^{\dag}b_{k-q,\sigma}a_{p+q,-\sigma}
- a_{k\sigma}^{\dag}b_{p,-\sigma}^{\dag}b_{k-q,-\sigma}a_{p+q,\sigma})$$  

\noindent Use of a usual BCS-like argument to derive SC from a static Hartree-Fock decoupling of $H_{res}^{(2)}$ is possible~\cite{TMDSC}, but problematic.  This is because, as it stands, the effective interaction leads to possibility of both p-p and p-h condensation, and at intermediate-to-strong ''excitonic Kondo'' coupling, it is known that there is strong interference between both these channels at two-particle level.  Then the associated logarithmic divergences appear in both channels, best illustrated within 
the classic parquet approach~\cite{abrikosov}.  We have adapted Abrikosov's original parquet approach to our case (see SI for details).  We proceed as follows: the bare vertex, $\Gamma_{\sigma_{1}\sigma_{2}\sigma_{3}\sigma_{4}}^{(0)}(p_{1},p_{2},p_{3},p_{4})=J(\delta_{\sigma_{1}\sigma_{3}}\delta_{\sigma_{2}\sigma_{4}}-\delta_{\sigma_{1}\sigma_{4}}\delta_{\sigma_{2}\sigma_{3}})$ will obtain drastic renormalization in the `excitonic Kondo' regime. To this end, we exploit DMFT results: since the $a$-band spectral function is gapped, one 
can replace $G_{aa}^{-1}(\omega)\simeq \omega +\epsilon_{a}$ at low energy 
(since Im$\Sigma_{aa}(\omega)=0$ at low energy), while the $b$-band propagator 
is approximated as $G_{bb}^{-1}(k,\omega)\simeq \omega -z_{b}\epsilon_{k,b}$, 
with $z_{b}^{-1}=1-(d/d\omega)$Re$\Sigma_{bb}(\omega)|_{\omega=E_{F}}$.  
With these inputs, we find that the vertex is logarithmically enhanced, giving 
$\Gamma(\omega)=J[1+\rho$ln$(E_{F}/|\omega|+\epsilon_{a})]^{-1}$.  
The SC transition temperature is estimated from the divergence of the 
renormalized vertex, and we find $T_{c}=E_{F}.e^{-1/J\rho}$.  
Using the {\it renormalized} $E_{F}\simeq T_{K}\simeq 100$~K from DMFT 
results below which correlated FL behavior sets in (instead of the {\it bare} 
$E_{F}\simeq 23$~meV) in $Bi$, $\rho(E_{F})\simeq 0.1$eV$^{-1}$ from normal state DMFT results, and the 
coupling $J\simeq O(1)$ (we are {\it not} in the regime $t_{11,22,12}<<U'$ for 
$Bi$), we estimate the SC $T_{c}\simeq O(1)$~mK, quite close to the 
experimental finding of $T_{c}\simeq 0.5$~mK.  Given the approximations 
made above, this is very reasonable.  Our estimate is a huge enhancement compared to 
the $T_{c}^{BCS}\simeq O(100)$~nK found using the naive
BCS formula~\cite{ramky}, and reflects the strong coupling nature of SC, 
where non-adiabatic effects enhance $T_{c}$ via huge enhancement of the vertex.  Effects akin to the above have been discussed in the non-adiabatic limit of strong electron-phonon coupling~\cite{pietronero} and, in $Bi$,
the fact that $g/E_{F}\simeq O(0.5)$ will also imply such additional enhancement arising from electron-phonon 
coupling.  In the non-adiabatic limit, it is possible to subsume this latter effect into a renormalized value 
of the bare vertex $J$ and, in fact, our DMFT results do include the renormalization caused by e-p coupling as well. Such strong coupling multi-band SC will also generally give an enhanced $(dH_{c2}/dT)$ in $Bi$~\cite{ramky}, as in other documented near semi-metallic superconductors~\cite{singh}, relative to the BCS prediction.

\noindent Finally, the form of the effective interaction also shows that the excitonic 
insulator (EI) phase is a subleading instability in $Bi$.  It is of interest to 
inquire whether evidence for such a ``competing order'' could obtain in $Bi$ 
by modification of its electronic structure, $e.g$, by pressure~\cite{Bi-MIT} 
or an external magnetic field.  Both reduce the tiny carrier density further, 
jacking up the effective $U/t$ ratio and inducing tendency to localization.
Actually, pressure does induce a metal-insulator-like transition in 
Bismuth~\cite{Bi-MIT}.  Within our strong coupling analysis, 
the competing order in the resultant semiconducting phase would be characterized by an order parameter 
$\Delta_{ex}=\sum_{k,\sigma}\langle f_{12}(k)c_{1k\sigma}^{\dag}c_{2k\sigma}\rangle$ with $f_{12}(k)=2\sqrt{3}$sin$(\sqrt{3}k_{x}a/2)$sin$(k_{y}a/2)$.  This excitonic order parameter does not break inversion symmetry
(${\bf k}\rightarrow -{\bf k}$) but, remarkably, is associated with an electronic {\it nematic} order~\cite{CKL}. Very recent work~\cite{yazdani} finds a nematic electronic state with a tiny gap $O(500)\mu$eV on the surface of $Bi$ under high magnetic fields. Whether a Rashba-SOC modified electronic structure as above can induce an excitonic instability as proposed here, and whether such a state can lead to a reconstructed electronic structure having the observed anisotropy of Landau level wave-functions, is an enticing open issue.  Thus, future studies under pressure can confirm or refute our prediction which, at this time seems to have limited confirmation~\cite{yazdani}. Investigation into these aspects is left for future work.

\vspace{1.4cm}

\noindent {\bf Supplementary Information}
   Here, we adapt the parquet approach of Bychkov {\it et al.}~\cite{abrikosov} to our model. 
  In the parquet approach~\cite{abrikosov}, graphs corresponding to p-h and p-p vertices which cannot be cut into two separate pieces by cutting two ($a$, the ``heavy'' band, or $b$, the metallic band in $Bi$) propagator lines are neglected.  However, the leading logarithmic corrections arising from Kondo screening are retained.  At $N$th order, the 
magnitude of each
diagram for $\Gamma$is $x^{N}$, where $x=J\rho\int G_{aa}G_{bb} d\omega d\epsilon\simeq J\rho$ln$(E_{F}/\omega+\epsilon_{a})\simeq O(1)$.  The full vertex in the parquet approximation is $\Gamma=\Gamma_{0} + \Lambda_{pp} + \Lambda_{ph}$,
with $\Gamma_{0}$ the bare vertex, $\Lambda_{pp}$ the p-p ``brick'' which can be cut by two parallel $a,b$ propagator lines, and $\Lambda_{ph}$
 the corresponding brick in the p-h channel.  The parquet series is analyzed by requiring $(i)$ no 
energy transfer to and from the internal (gapped) $a$-band (gapped in case of 
$Bi$, see DMFT results) states; thus, one can set $\omega_{2}=\omega_{3}=0$ and $\omega_{1}=\omega_{4}=\omega$, $(ii)$ all momenta $p_{i}, i=1,2,3,4$ are set 
equal to $p_{F}$.  $(iii)$ using $\Gamma(\omega,0,0,\omega)=\Gamma(\omega)$, 
and choosing an {\it internal} two-line state such that the energy $\omega'$ of 
the $G_{aa}$ line is minimum.  Since there is a {\it full} vertex part $\Gamma(\omega')$ to the left and right of this part, 
one finds~\cite{abrikosov},

\be
\Gamma_{\sigma_{i}}^{P}(\omega)=\rho\int_{|\omega|}^{E_{F}}\frac{\Gamma_{\sigma_{1}\sigma_{4}\mu\nu}(\omega')\Gamma_{\mu\nu\sigma_{2}\sigma_{3}}(\omega')}{\omega'}d\omega'
\ee
and

\begin{equation}
\Gamma_{\sigma_{i}}(\omega)=\rho\int_{|\omega|}^{E_{F}}\frac{\Gamma_{\sigma_{1}\nu\mu\sigma_{4}}(\omega')\Gamma_{\mu\sigma_{2}\sigma_{3}\nu}(\omega')}{\omega'}d\omega'
\end{equation}
  Using the spin-separable structure of $\Gamma$, one obtains 
$\Gamma(x)=J - J\rho\int_{0}^{x}\Gamma^{2}(x') dx'$, 
with $x=$ln$(E_{F}/|\omega|+\epsilon_{a})$.  
This has the solution 
$\Gamma(\omega)=J[1+\rho$ln$(E_{F}/|\omega|+\epsilon_{a})]^{-1}$, 
showing up the logarithmic enhancement of the two-particle vertex.  
Hence, with $J >0$ (notice that this co-efficient of 
$H_{res}^{(2)}$ is {\it negative} as above), $\Gamma(\omega)$ 
has a pole when $\Gamma^{-1}(\omega_{c}=T_{c})=0$, signalling an instability 
to the superconducting state when
\begin{figure}[h]
\includegraphics[angle=270,width=\columnwidth]{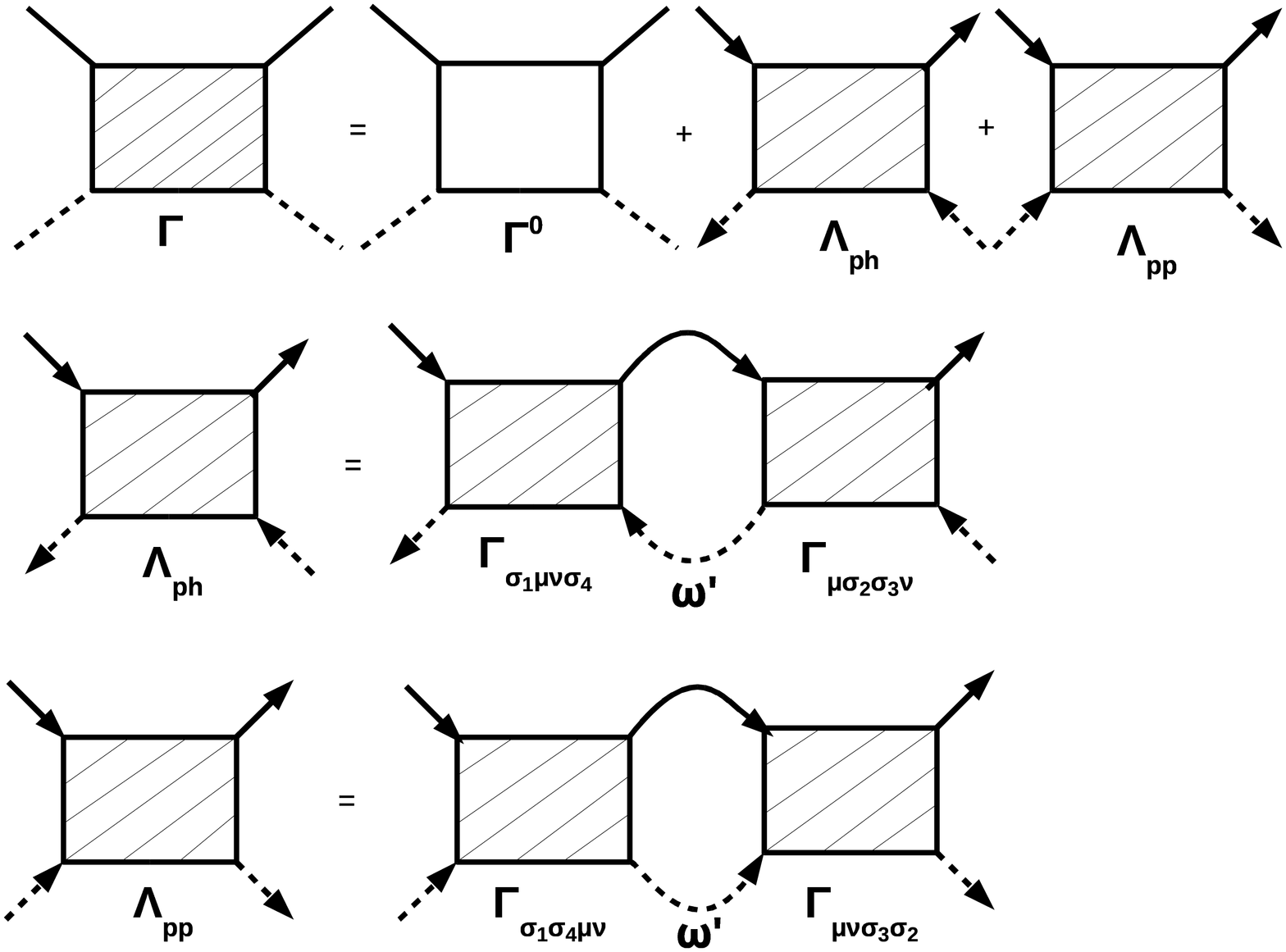}
\caption{The diagrammatic representation of the parquet approximation for the full vertex function (upper figure) and the blocks in the particle-hole (ph, center figure) and particle-particle (pp, lower figure) channels (see text for details).}
\label{fig5}
\end{figure}
$T_{c}\simeq E_{F}.e^{-1/J\rho}$.  Thus, this enhancement of $T_{c}$ relative 
to the simple weak-coupling BCS estimate is a consequence of the logarithmic 
enhancement o the vertex function.  Using the {\it renormalized} 
$E_{F}\simeq T_{K}\simeq 100$~K from DMFT results below which correlated FL behavior sets in (instead of the {\it bare} $E_{F}\simeq 23$~meV in $Bi$), 
$\rho(E_{F}\simeq 0.1$eV$^{-1}$, and the coupling $J\simeq O(1)$ (we are 
{\it not} in the regime $t_{11,22,12}<<U'$ for $Bi$), we estimate the SC 
$T_{c}\simeq O(1)$~mK, quite close to the experimental finding of 
$T_{c}\simeq 0.5$~mK.  

   The above parquet analysis also suggests that $T_{c}$ could be enhanced further if it would be possible to move the $a$-band states 
($\epsilon_{a}$) closer to the Fermi energy by appropriate perturbations.

\vspace{1.4cm}
\bibliographystyle{apsrev4-1}

\begin{thebibliography}{60}
\bibitem{Bi-MIT} N. P. Armitage, R. Tediosi, F. L\'evy, E. Giannini, L. Forro, and D. vander Marel, {\it Phys. Rev. Lett.} {\bf 104}, 237401 (2010).
\bibitem{behnia} K. Behnia, L. Balicas, and Y. Kopelevich, {\it Science} {\bf 317}, 
1729 (2007).
\bibitem{ramky} Om Prakash, A. Kumar, A. Thamizhavel and S. Ramakrishnan, 
arXiv:1603.04310, to appear in {\it Science}.

\bibitem{giamarchi} P. Chudzinskii and T. Giamarchi, {\it Phys. Rev. B} {\bf 84}, 125105 (2011).

\bibitem{armitage} R. Tediosi, N. Armitage, E. Giannini, and D. vander Marel, 
{\it Phys. Rev. Lett.} {\bf 99}, 016406 (2007).

\bibitem{ting1994} J. H. Xu, E. G. Wang, C. S. Ting, and W. P. Su, {\it Phys. Rev. B} {\bf 48}, 17271 (1993).

\bibitem{cohen} M. L. Cohen, {\it Phys. Rev.} {\bf 134}, A511 (1964).

\bibitem{2H-TaSe2} A. Taraphder, S. Koley, N. S. Vidhyadhiraja and M. S. Laad, 
{\it Phys. Rev. Lett.} {\bf 106}, 236405 (2011).
\bibitem{kukkonen-J.Phys.F-1977}C. A. Kukkonen and K. F. Sohn Journal of Physics F: Metal Physics, {\bf 7}, L193 (1977). 
\bibitem{1T-TiSe2} S. Koley, M. S. Laad, N. S. Vidhyadhiraja and A. Taraphder, {\it Phys. Rev. B} {\bf 90}, 115146 (2014); S. Koley, N. Mohanta and A. Taraphder, {\it J. Phys. Condens. Matter} {\bf 27} 185601 (2015). 
\bibitem{jarrell} N. Dasari, W. R. Mondal, P. Zhang, J. Moreno, M. Jarrell
and N. Vidhyadhiraja, {\it Eur. Phys. Journal B}, (in press) (2016).
\bibitem{U<0Hubbardmodel} M. Keller, W. Metzner, and U. Schollwöck, {\it Phys. Rev. Lett.} {\bf 86}, 4612 (2001).
\bibitem{mohit} S. S. Kancharla and S. Okamoto, {\it Phys. Rev. B} {\bf 75}, 193103 (2007). 
\bibitem{laad-122} S. D. Das, M. S. Laad, L. Craco, J. Gillett, V. Tripathi, and S. E. Sebastian, {\it Phys. Rev. B} {\bf 92}, 155112 (2015); L. de' Medici, 
S.R. Hassan, M. Capone, Xi Dai, {\it Phys. Rev. Lett.} {\bf 102},126401 (2009).
\bibitem{biermann} J. Tomczak and S. Biermann, {\it Phys. Rev. B} {\bf 80}, 085117 (2009). 
\bibitem{CKL} A. Taraphder and P. Coleman, {\it Phys. Rev. Lett.} {\bf 66}, 2814 (1991).
\bibitem{TMDSC} S. Koley, arXiv:1606.02841.
\bibitem{abrikosov} Yu. Bychkov, L. Gorkov and I. Dzyaloshinskii, {\it J. Exptl. Theor. Phys. (U.S.S.R)} {\bf 50}, 738 (1966); K. Svozil, Physica Status Solidi, {\bf 147}, 635 (1988). 
\bibitem{pietronero}  L. Pietronero1 and S. Straessler, {\it Europhys. Lett.} {\bf 18}, 627 (1992).
\bibitem{singh}  DJ Singh,  {\it PLoS ONE} {\bf 10}, e0123667 (2015). 
\bibitem{yazdani} Benjamin E. Feldman, Mallika T. Randeria, András Gyenis, Fengcheng Wu, Huiwen Ji, R. J. Cava, Allan H. MacDonald, Ali Yazdani, arXiv:1610.07613 (2016).
\end{thebibliography}

\end{document}